\documentclass[twocolumn,aps,prl,floats,superscriptaddress]{revtex4-1}
\usepackage[T1]{fontenc}
\usepackage[utf8]{inputenc}
\usepackage{amsmath}
\usepackage{graphicx}
\usepackage{amssymb}
\usepackage{ifthen}

\usepackage{amssymb,amsmath}
\usepackage[usenames]{color}
\usepackage{graphicx}
\newcommand{\op}[1]{\widehat{#1}}
\newcommand{\dagop}[1]{\widehat{#1}^{\dagger}}
\newcommand{\bo}[1]{{\mathbf{#1}}}

\newcommand{\wb}[1]{{\overline{#1}}}

\newcommand{\etal}{~\textsl{et al.}}

\newlength{\templength}

\newcommand{\eqn}[1]{(\ref{#1})}

\newcommand{\eq}[2]{\begin{equation}\label{#1}#2\end{equation}}

\newcommand{\out}[1]{} 							%uncomment this to remove "deleted" text in a length check

\newcommand{\REM}[1]{\ifthenelse{0=1}{#1}{}}

\begin{document}

\title{Correlation evolution in dilute Bose-Einstein condensates after quantum quenches}

\author{J. Pietraszewicz}
\email{pietras@ifpan.edu.pl}
\affiliation{Institute of Physics, Polish Academy of Sciences, Al. Lotnik\'ow 32/46, 02-668 Warsaw, Poland}
 
\author{M. Stobi\'nska}
\altaffiliation[Present address: ]{University of Warsaw}
\affiliation{Institute of Physics, Polish Academy of Sciences, Al. Lotnik\'ow 32/46, 02-668 Warsaw, Poland}
\affiliation{Faculty of Physics, University of Warsaw, ul. Pasteura 5, 02-093 Warsaw, Poland}

\author{P. Deuar}
\email{deuar@ifpan.edu.pl}
\affiliation{Institute of Physics, Polish Academy of Sciences, Al. Lotnik\'ow 32/46, 02-668 Warsaw, Poland}

\date{\today }
\begin{abstract}
The universal forms of quantum density and phase correlations after an interaction quench are found for dilute 1d, 2d, and 3d condensates. A Bogoliubov approach in a local density aproximation is used. We obtain compact expressions for the most visible effects. Our results show how loss of phase coherence and antibunching are built up after the quench by quantum fluctuations. 
 We demonstrate further that the density correlations can be observed even with imaging resolution much worse than healing length. 
This indicates that the direct measurement of counterpropagating atom pairs \emph{in situ} in a continuum system is realistic. The conditions in contemporary 1d experiments are especially favorable for the correlation wave observations.  
\end{abstract}
%\pacs{}

\maketitle

%%%%%%%%%%%%%INTRO%%%%%%%%%%%%%%%%%%%%%%%%%%%%%%
\label{INTRO}
Quantum quenches are one of the fundamental quantum dynamical phenomena in many-body systems and cosmology  \cite{Langen15a,Essler16,Mitra18}.
They are generically induced by changes to the Hamiltonian that occur globally and non-adiabatically, and become visible when the relevant energy scale rises above the thermal one.
Cold atom systems have enabled the investigation of this kind of non-equilibrium quantum dynamics to an unprecedented degree. 
The greatest emphasis has been placed on strong quenches across a phase transition, and production of entanglement and complexity. Examples, often on lattice systems, include 
 \cite{Kormos14,Cheneau12,Trotzky12,Kollath07,Alba18,Fitzpatrick18,Fischer08}. 

Quenches that remain within the condensate phase of dilute continuum systems are also of much interest, though less widely explored. A major motivation to understand them is that they occur in many existing experiments. Continuum quenches are interesting also because the Lieb-Robinson bound  \cite{LiebRobinson72} does not apply to the excitations since hopping speed is not limited by a lattice. A quantum quench can be wilfully imposed by varying the tight confinement or the interaction strength.  
A quench can also  happend as a side effect: when preparing the initial state of reduced-dimensional gases, or from a rapid loss of atoms that affects the chemical potential. Time-varying spatial correlations are induced, and indeed have been measured in contemporary cold atom experiments  \cite{Langen13,Hung13,Cheneau12,Trotzky12} 

Research on quenches in dilute continuum condensates include the comprehensive work of Calabrese, Caux, and Cardy
 on the dynamical structure factor after a quench \cite{Calabrese06,Calabrese07a,Calabrese07b},
  the Lieb-Liniger 1d model \cite{Caux06,Caux07,Zill15,Zill16}, and its long-time steady state \cite{DeNardis14}.
 Density correlations and waves in a 3d BEC were studied by Carusotto\etal  \cite{Carusotto10}.
 The equilibration of phase correlations in 1d was measured in \cite{Langen13}, and density structure factors have also been measured  in 2d dilute gases \cite{Hung11,Rancon13}.
Recently, an extensive study of phase coherence and momentum distributions has been made \cite{Martone18}.

\begin{figure}
\includegraphics[width=\columnwidth]{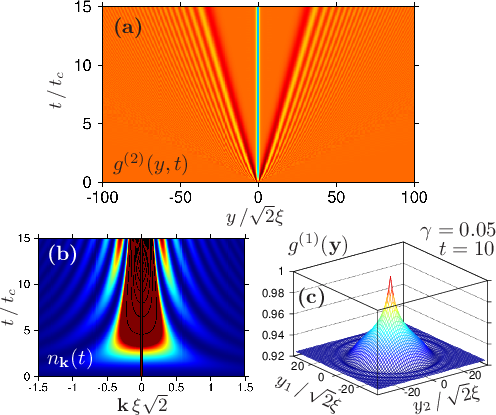}\vspace*{-1em}
\caption{Correlations after a quantum quench from $g_0\sim0$ to $g\gg g_0$:
 (a) density correlation\ $g^{(2)}(y,t)$ and
 (b) momentum distribution\ $n_{\bo{k}}(t)$ in a 1d gas, and
 (c) a snapshot of the phase correlation\ $g^{(1)}(\bo{y})$ in 2d.  
 Red/blue color indicates high/low values. Color brown in (b) is a saturated high value, with black higher contours superposed.
 The healing length $\xi$, $\gamma$ parameter, distance $y$, and time unit $t_c=\hbar/mc^2$ are as in the text. 
\label{fig1}}\vspace*{-1em}
\end{figure}

Compact analytic expressions for the quench correlations in dilute gases have not been available. It is the
purpose of this paper to fill that gap. We will study
a quench of the contact interaction at zero temperature, and consider uniform sections of a gas in the Bogoliubov approximation.
We obtain expressions for density and phase correlations across the whole gamut of dimensionalities and
post-quench times. They go beyond the two earlier studies of particular cases  \cite{Carusotto10,DeNardis14}.
This includes medium times in 1d, which we find to be especially favorable for observations with realistic detector resolution.

The resulting correlations have length scales comparable to the healing length, much shorter than the typical size of the condensate.
Therefore, what we obtain can be fed into a local density approximation (LDA) to describe most non-uniform cases of interest. The first and second-order spatial correlation functions, $g^{(1)}$ and $g^{(2)}$, give an intuitive picture of the behavior that occurs in single
realizations of the gas. The shape of the correlations
matches the shape of typical disturbances in the gas.
Representative examples are shown in Fig.~\ref{fig1}.

%%%%%%%%%%%%%SYSTEM%%%%%%%%%%%%%%%%%%%%%%%%%%%%%%

Consider a uniform $d$-dimensional Bose gas with
  contact interactions of strength $g$ and mean density $\wb{n}$. The Hamiltonian, in terms of a Bose field $\op{\Psi}(\bo{x})$, is: 
\eq{HPsi}{
\op{H} = \int\, d^d\bo{x}\ \dagop{\Psi}(\bo{x})\left[
-\frac{\hbar^2}{2m}\nabla^2 + \frac{g}{2}\, \dagop{\Psi}(\bo{x})\op{\Psi}(\bo{x})
\right]\op{\Psi}(\bo{x}).
}
 The system  has just one dimensionless parameter  
\eq{gamma}{
\gamma = \left(\frac{m\, g\, \wb{n}}{\hbar^2}\right)^d \ \frac{1}{\wb{n}^{2}}.
} 
 It is self-same with the Lieb-Liniger gamma parameter in the 1d gas \cite{Lieb63}, and with\ $\gamma=(4\pi)^3\, \wb{n}a_s^3$\ in 3d, 
 where $a_s$ is $s$-wave scattering length.
 In reduced dimensionalities, the confinement in the tightly bound directions affects\ $\gamma$, e.g.   
 in 1d $g = 4\pi\hbar a_s\nu_{\perp}$, where $\nu_{\perp}$ is the tight trapping frequency. 
 For our purposes we introduce length  $u_x=\hbar/mc$ and time  units  $t_c = mu_x^2/\hbar$, with $c=\sqrt{g\wb{n}/m}$ 
 the speed of sound and $\xi=\hbar/\sqrt{2mg\wb{n}}$ the healing length. 
Setting $\hbar=m=c=1$ one obtains dimensionless variables with
$\xi =  \frac{1}{\sqrt{2}}$ and $\gamma = 1/\wb{n}^2$. 

 Let us study the paradigmatic quench from the initially pure non-interacting condensate
 in the zero momentum ($\bo{k}=0$) mode.
 Instantaneously at $t=0$, the interaction is turned on to its final value $\gamma~>~0$.  
The evolution is treated using a standard number-conserving Bogoliubov description \cite{Castin98}, with similarities to \cite{Carusotto10}. 
 The Bogoliubov approach boils down to two assumptions: 
 (I)\ the quantum depletion $\delta N/N$, being the fraction of atoms in the non-condensate modes,  
 is much smaller than one, and\ (II)\ the interaction between non-condensate modes can be neglected. 
Assumption (II) is met when $\gamma\ll1$, while (I) requires that the volume $V$ is mostly phase coherent. In 3d this is easily met, while in 2d and 1d it restricts the allowed box size $V$ or time.
The calculation provides values of $\delta N/N$ and $g^{(1)}(y)$ so that validity can be checked self-consistently. 

To proceed, space of volume $V=L^d$ can be discretized on an arbitrarily fine lattice with volume $\Delta v$ per point $\bo{r}_i$. The Hamiltonian (1) becomes then:
\eq{HBH}{
\op{H} = -J \sum_{i,j} \dagop{a}_{i}\op{a}_{j} + \frac{U}{2}\sum_i\dagop{a}_i\dagop{a}_i\op{a}_i\op{a}_i,
}
in terms of creation $\dagop{a}_i$ and annihilation $\op{a}_i$ operators on the lattice with $J=1/[2(\Delta v)^{2/d}]$,  $U=\sqrt{\gamma}/(\Delta v)$, and mean site occupation $\Delta v/\sqrt{\gamma}$. 
Since a distance scale of at least the healing length is required to encompass the continuum physics, we need $\Delta v\ll1$.  
 This way one assures that mapping of the continuum onto the lattice is sensible.  
 On a square lattice, $\Delta v$ corresponds to a maximum momentum cutoff $k_{\rm max} = \pi/(\Delta v)^{1/d}$,
 and so $J/U = \frac{1}{2\sqrt{\gamma}} \ (\frac{k_{\rm max}}{\pi})^{2-d}$.

%%%%%%%%%%%%%MAIN RESULTS%%%%%%%%%%%%%%%%%%%%%%%%%%%
 
We find the phase coherence and density correlations
 at a distance $y = |\bo{y}|$ between two points $\bo{r}$ and $\bo{r}'=\bo{r}+\bo{y}$, 
 in terms of the dispersion relation for $k=|\bo{k}|$:
$\omega_k~=~k~\sqrt{1+k^2/4}$.
 The normalized phase coherence is given by
\begin{eqnarray}\label{g1def}
\lefteqn{
\!\!\!g^{(1)}(y,t) =\frac{\langle\dagop{a}(\bo{r})\op{a}(\bo{r}')\rangle}{\wb{n}\Delta v}= 1 - \frac{1}{2\wb{n}V}\sum_{\bo{k}\neq0}\frac{1}{\omega_k^2}\qquad}&\nonumber\\
&\times \left[1-\cos2\omega_kt-\cos \bo{k}\cdot\bo{y} + \cos(\bo{k}\cdot\bo{y}+2\omega_kt)\right].\ 
\end{eqnarray}
The normalized density correlations are given by 
\begin{eqnarray}\label{g2def}
\lefteqn{
g^{(2)}(y,t) = \frac{\langle\dagop{a}(\bo{r})\dagop{a}(\bo{r}')\op{a}(\bo{r}')\op{a}(\bo{r})\rangle}{(\wb{n}\Delta v)^2}
} &\nonumber\\
& =1 - {\displaystyle\frac{1}{2\wb{n}V}}{\displaystyle\sum\limits_{\bo{k}\neq0}}{\displaystyle\frac{k^2}{\omega_k^2}}\,\left[\cos \bo{k}\cdot\bo{y} - \cos(\bo{k}\cdot\bo{y}+2\omega_kt)\right].\quad
\end{eqnarray}
 We find also the mode occupation to be
\begin{eqnarray}\label{nk}
n_{\bo{k}}~=~\left[(\sin \omega_kt)/\omega_k\right]^2,
\end{eqnarray}
which appears in the quantum depletion   
\begin{eqnarray}
\delta N(t)/N = \frac{1}{N}\sum_{\bo{k}\neq0}n_{\bo{k}}.
\end{eqnarray}

 In the large system and continuum limit ($V~\to~\infty$, $\Delta v\to~0$)
 the discrete sums over $\bo{k}$ can be converted to integrals. 
 Then, for each dimensionality\ $d$, one has the solutions 
\begin{eqnarray}\label{g1int}
g^{(1)}(y,t) &=& 1 - \sqrt{\gamma}\int_0^{\infty}\hspace*{-3ex}\, dk\ \left(\frac{1-\cos 2\omega_k t}{k^2+4}\right) 
\frac{1-M_d}{a_d\,k^{3-d}}\qquad \\
\label{g2int}
g^{(2)}(y,t) &=& 1 - \sqrt{\gamma}\int_0^{\infty}\hspace*{-3ex}\, dk\ \left(\frac{1-\cos 2\omega_k t}{k^2+4}\right) 
\frac{M_d\,k^{d-1}}{a_d}\qquad 
\end{eqnarray}
 as a function of time\ $t$ and distance~ $y$.
 The functions $M_d=M_d(k\, y)$ and constants $a_d$ are the following
\eq{af}{
\hspace*{-0.7em}M_d = \left\{ 
\begin{array}{c@{\ }c}
\cos k y 			&	\text{for }d=1\\
J_0\left[k|y|\right] 	&	\text{for }d=2,\\
\frac{\sin k y}{k y} 	&	\text{for }d=3
\end{array}
\right.
\, 
a_d = \left\{ \begin{array}{c@{\ }c}
\pi/2 			&	\text{for }d=1\\
\pi 	&	\text{for }d=2\\
\pi^2 	&	\text{for }d=3 
\end{array}\right.\!\! 
}
and $J_{\alpha}\left[x\right]$ are Bessel J functions.
 Notably, the solutions retain the same universal shapes, and only the deviations from full coherence $g^{(\mu)}=1$ are proportional to $\sqrt{\gamma}$. 
 The expressions are accurate, as long as the deviations are perturbative.
 Characteristic examples of \eqn{g1int}-\eqn{g2int} are shown in Fig.~\ref{fig1}.

%%%%%%%%%%%%%SPECIFIC CASES%%%%%%%%%%%%%%%%%%%%%%%%%

\begin{figure}[t]
\includegraphics[width=\columnwidth]{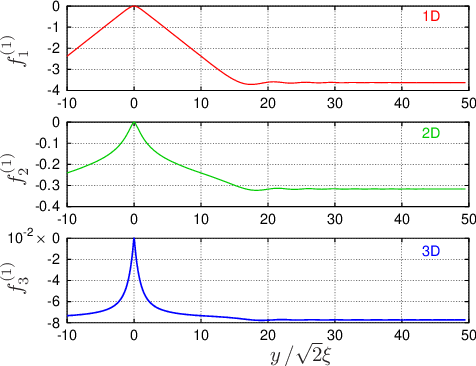}\vspace*{-1em}
\caption{Universal solutions (\ref{g1int}) for phase coherence\ 
 scaled as $f^{(1)}_d = [g^{(1)}(y,t)-1]/\sqrt{\gamma}$, \, shown at long times $t=7.5t_c$.
\label{figg1}}
\end{figure}

\begin{figure}[b]
\includegraphics[width=\columnwidth]{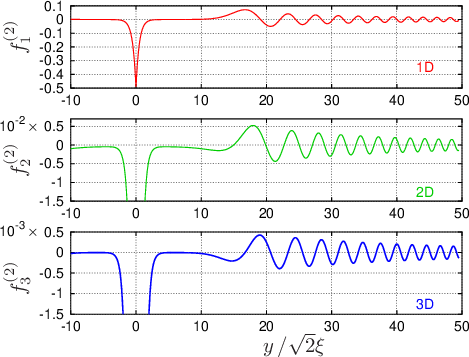}\vspace*{-1em}
\caption{Universal solutions (\ref{g2int}) for density correlations\  scaled as $f^{(2)}_d = [g^{(2)}(y,t)-1]/\sqrt{\gamma}$,\  shown at long times $t=7.5t_c$.
\label{figg2}}
\end{figure}

 Explicit expressions for the correlations \eqn{g1int}-\eqn{g2int}
 can be found in various limits.  
 Fig.~\ref{fig1}(a)  depicts the two principal regimes:
 a timelike regime inside the sound cone, and a spacelike one outside. 
 Between them, the main correlation wavepacket is propagating with twice the speed of sound. 
 The double speed comes about because it is composed of counterpropagating atom pairs.
 Representative profiles  of the correlations  at a given time $t=7.5t_c$
 are shown in more detail in Figs.~\ref{figg1} and~\ref{figg2}. 

 In the \emph{spacelike regime} $y>2t$, at long times~ $t~\gtrsim~1$,
 the density fluctuations decay to $g_{\rm spacelike}^{(2)}=1$. 
 The~ phase coherence 
 \eq{g1sp}{
g^{(1)}_{\rm spacelike}(y,t) \approx 1-\delta N(t)/N,
}
is reduced by the depletion
\eq{depl}{
\frac{\delta N(t)}{N} \approx \sqrt{\gamma}\times\left\{\begin{array}{c@{\quad}l}
\frac{4t-1}{8}								&\text{for }d=1\\
\frac{1}{4\pi}\left(\ c_1 + \log t\ \right)					&\text{for }d=2\\
\frac{1}{4\pi}\left(1 + c_2 e^{-c_3 t}\right)	&\text{for }d=3.
\end{array}\right.
}
 The main contributions to this reduction at long times come from phonon excitations. 
 The constants $c_j$ are found numerically, and are $c_1\approx1.96(1)$, $c_2\approx0.43(1)$, and $c_3\approx2.24(2)$. 
 We see the linear and logarithmic decay characteristic of 1d and 2d quasicondensates, respectively.
 Clearly,  in this regime the model is breaking down  at long enought times, when $\delta N/N\to\approx 1$,  but remains predictive for shorter ones.
 In the 3d case, the loss of long-range coherence stabilizes at
 a constant value, as expected for a BEC.
 However, this value $\delta N/N=\frac{\sqrt{\gamma}}{4\pi}\approx0.080\sqrt{\gamma}$, seen also in \cite{Martone18},  
is larger than the ground state value of\  $\frac{\sqrt{\gamma}}{3\pi^2}\approx0.034\sqrt{\gamma}$ \cite{Dalfovo99}.
 This indicates  that the quench excites many more pairs
 than appear in the ground state. 

 In the \emph{timelike regime} \ $y<2t$, and at long times~  $t~\gg~1$, the phase coherence:  
\eq{g1tim}{
\hspace*{-1ex}g^{(1)}_{\rm timelike}(y) \approx 1 - \sqrt{\gamma}\times \left\{\begin{array}{c@{\ }l}
\frac{2y-1}{8}								&\text{for }d=1\\
\frac{1}{4\pi}\left(\ \gamma_E + \log y \ \right)			&\text{for }d=2\\
\frac{1}{4\pi}\left(1-\frac{1}{2y}\right)									&\text{for }d=3
\end{array}\right.
}
 decays for $y\gtrsim 1$  to its spacelike value \eqn{g1sp}, where $\gamma_E\approx~0.5772$.   
The validity of the description breaks down, when the deviation of $g^{(1)}$ becomes comparable with unity.  This  occurs   for distances $y\gtrsim\frac{1}{\sqrt{\gamma}}$ in 1d, and $y\gtrsim\exp(\pi/\sqrt{\gamma})$ in 2d. 

The long times density correlations  
\eq{g2tim}{
g^{(2)}_{\rm timelike}(y) \approx 1 - \sqrt{\gamma}\times\left\{\begin{array}{c@{\ }l}
\frac{e^{-2y}}{2}\vspace*{0.3em}								&\text{for }d=1\\
\frac{K_0\left[2y\right]}{\pi}\vspace*{0.3em}						&\text{for }d=2\\
\frac{e^{-2y}}{2\pi y}									&\text{for }d=3
\end{array}\right.
}
exhibit anti-bunching on a healing-length scale.  	
 $K_0\left[x\right]$ in (\ref{g2tim}) is the Bessel K function. 

 The state reached in \eqn{g2tim} differs from the 1d ground state, which has   
antibunching of $g^{(2)}(0)=1-\frac{2}{\pi}\sqrt{\gamma}\approx 1-0.637\sqrt{\gamma}$ \cite{Kheruntsyan03}. 
 However, it does agree with earlier studies \cite{DeNardis14} that found the stationary state to be
 a~ peculiar one, i.e. neither thermal nor described by a generalized Gibbs ensemble. 
 For $d\ge2$, the expressions~ \eqn{g2tim} are divergent at $y=0$. In practice, the~ dip in correlation 
 is limited to a value determined  
by the accessible momentum:
\eq{g2kmax}{
g^{(2)}_{\rm timelike}(0) = 1 - \sqrt{\gamma}\times\left\{\begin{array}{c@{\ }l}
\frac{1}{\pi}\tan^{-1}\frac{k_{\rm max}}{2}\vspace*{0.3em}				&\text{for }d=1\\
\frac{1}{2\pi}\log\left(1+\frac{k_{\rm max}^2}{4}\right)\vspace*{0.3em}						&\text{for }d=2\\
\frac{1}{\pi^2}\left(k_{\rm max}-2\tan^{-1}\frac{k_{\rm max}}{2}\right)									&\text{for }d=3
\end{array}\right.
}

 The time dependence of $g^{(2)}(0)$ quantifies the onset of antibunching.
 It can be useful to judge prethermalization timescales. 
 One finds that in 1d, the form
\eq{g20tim}{
g^{(2)}(0) \approx 1 - \sqrt{\gamma}\times\left\{\begin{array}{c@{\ }l}
\frac{3}{2}\sqrt{t} -\frac{1}{2}t^{3/2}								&\text{for }t\lesssim \tfrac{1}{4}\\
\frac{1}{\pi}\tan^{-1}\frac{k_{\rm max}}{2}-\dfrac{c_2e^{-c_3t}}{2t^{c_4}}			&\text{for }t\gtrsim \tfrac{1}{4}
\end{array}\right.
}
fits the calculated function quite well.   
The constants are $c_2=0.35(1)$, $c_3=2.05(1)$ and $c_4=0.33(2)$.

Importantly, 
the bulk of the timelike regime is uncorrelated because no disturbance can travel slower than the speed of sound. 

On the boundary between the spacelike and timelike regimes, at $y\approx2t$, one find the main density correlation wave.
It takes the forms:
\eq{g2wave}{
\hspace*{-1em}g^{(2)}_{\rm wave}(y) \approx 1 + \sqrt{\gamma}\times\!\left\{\!\!\begin{array}{c@{\,}l}
\frac{1}{2\,(6t)^{1/3}}{\ \rm F_1}[-x]		&\text{for }d=1\\
\frac{1}{2\sqrt{\pi y}\,(6t)^{1/2}}\ F_2[-x]		&\text{for }d=2\!\\
\frac{1}{2\pi y\, (6t)^{2/3}}\ F_3[-x]				&\text{for }d=3
\end{array}\right.\!\!
}
where\ $x=(\frac{4}{3t})^{1/3}\,[y-2t]$\ is a scaled coordinate measuring the distance to the sound cone edge. 
\eq{Ffun}{
\hspace*{-1em}F_d[x] = \frac{1}{\pi}\int_0^{\infty}\!\!\!du\ u^{\frac{d-1}{2}}\cos\left[xu+\frac{u^3}{3} + \frac{\pi}{4}(d-1)\right]
}
 are generalizations of the Airy function Ai and for instance $F_1[x]~=~{\rm Ai}[x]$. 
 In the same boundary region as (\ref{g2wave}),
 the phase coherence $g^{(1)}$ has corresponding less visible oscillations.
They can be seen to the right of the kink in Fig.~\ref{figg1}, beyond the sound cone. This feature is also seen in \cite{Martone18}, but in earlier Luttinger liquid predictions \cite{Langen13} it was absent.
 
%%%%%%%%%%%%%INTRO%%%%%%%%%%%%%%%%%%%%%%%%%%%%%%%%%%%%%%%%%%%%%%%%%%%%%%%%%%%%%%%%%%%%%%%%%%%%%%%%%%%%%%%%%%%%%%%%%%%%%%%%%%%%%%%%%%%%%%
%{\sc  Observing the correlations:}\\

\begin{figure}
\includegraphics[width=0.75\columnwidth]{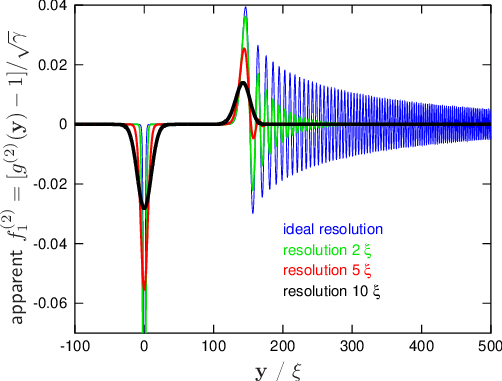}\vspace*{-1em}
\caption{
Density correlations $g^{(2)}(\bo{y})$ at $t=50t_c$ in 1d when seen with limited resolution. The number and amplitude of oscillations decrease as resolution worsens. From perfect (dark thin line), through $2\xi$, $5\xi$, to $10\xi$ (thick line).  
\label{fig4}}
\end{figure}

 Experiments, whether with absorption or phase-contrast imaging \cite{Gring12}, have resolutions of many healing lengths $\xi$. 
 Hence, the decay of coherence, $g^{(1)}$, inside the sound cone is directly resolvable,
 but not the antibunching dip or early correlation waves in $g^{(2)}$ (which must be measured \textsl{in situ}). 
 However, the correlation wave retains its structure in the \emph{scaled} length variable $x$,
 becoming magnified with time as $t^{1/3}$ and resolvable if the wave can survive intact for long enough. 
 Simultaneously, \out{from \eqn{g2wave},} its amplitude decays as $t^{-(2d-1)/3}$. 
 In Fig.~\ref{fig4} we emulate imperfect resolution measurements by convolving long-time $g^{(2)}(y)$
 with Gaussian point-spread functions. 
 One sees that the primary correlation wave and the anti bunching dip are robust to loss of resolution.
 They remain strongly visible and only mildly reduced
 in strength even with a (typical) resolution of $10\,\xi$.
 The high-velocity oscillations are rapidly lost. 
 The remaining disturbance resembels the compact correlation 
 wave seen with parity measurements in optical lattices \cite{Cheneau12}.
 Thus, finite imaging resolution leads to the appearance of an effective speed limit on what can be observed in the gas.
One may see similarity to a Lieb-Robinson bound, even though one does not formally apply here. 

To observe these waves in a simple manner,
 we should (A) have a gas that is long enough for the correlation wave to not meet the edge until it has broadened to a resolvable width,
 and (B) average over sufficiently many realizations for the correlation amplitude to emerge from the noise.

 Let us take a look at conditions in several
 1d gas experiments with ${}^{87}{\rm Rb}$.  
The Vienna experiment \cite{Gring12,Langen13} has had
$N\approx 900$ to $11 000$ atoms, trap frequencies\,  $\nu_x\times\nu_{\perp}\times\nu_{\perp} = 7\times1400\times1400$ Hz 
and observed the 1d phase correlation dynamics, similar to Fig.~\ref{figg1}.
They have an imaging resolution of $3.8\mu$m \cite{Smith11},  
 like in more recent work \cite{Erne18}.
 The  Palaiseau experiments   had  $N\approx1200$ atoms,  $4.5\mu$m resolution,
 trap frequencies $4\times3900\times3900$ Hz \cite{Armijo11}, or $7.5\times18800\times18800$~ Hz \cite{Jacqmin11}.

Now, think about the correlations  in the center of the cloud induced by the quench.
 The furthest they can cleanly propagate is a distance $y$ of about one Thomas-Fermi radius.
 For the lower atom number in the Vienna experiment, the time this takes is found to be $t\approx28t_c$ ($16$ms).  %\out{-this is for  density $n=20/\mu m$ } 
 The rms width of the main peak rising above $g^{(2)}=1$ is \, $w_{\rm rms}=1.802\times t^{1/3}$ 
 and the maximum height is $h_{\rm peak}=0.1474\times\sqrt{\gamma}/t^{1/3}$. These give $w_{\rm rms}=3.5\mu$m, and 
 i.e. $h_{\rm peak}=0.004$, respectively.
 The experimental resolution is in fact sufficient to resolve the structure even without the additional broadening seen in Fig.~\ref{fig4}. 
For the correlation peak to rise out of the shot noise, 
the statistical uncertainty should be less than  $h_{\rm peak}$. 
Taking a counting bin of the same size \, $w_{\rm rms}$ as the peak, with mean occupation~ $\wb{N}_{\rm bin}$, and shot noise ${\rm var}[N_{\rm bin}] = \wb{N}_{\rm bin}$, 
the variance of $g^{(2)}$ from one measurement is about $4{\rm var}[N_{\rm bin}]/\wb{N}_{\rm bin}^2$. Hence, the minimum number of realizations to average over 
is $4/(\wb{N}_{\rm bin}h_{\rm peak}^2)\approx 4000$. 
Clouds with higher atom numbers are generally less favorable both with regard to width in $\mu$m, which scales as $(N\nu_{\perp}\nu^4)^{-1/9}$, and the needed number of realizations, 
which scale as $(N^5\nu^2/\nu^4_{\perp})^{1/9}$.

 The Palaiseau experiment \cite{Armijo11} has more favorable conditions,
 with propagation time $t\approx80t_c$ ($28$ms), peak width $w_{\rm rms}=3.9\mu$m, 
 max.~ height $h_{\rm peak}\approx0.006$, and $2500$ required realizations.
 A~ comparison  with~ \cite{Jacqmin11}, which had higher\,  $\gamma\approx0.17$, \,  is instructive:
 $t~\approx~190t_c$ (15ms), peak width $w_{\rm rms}\approx2.5\mu$m,
 max. height $h_{\rm peak}=0.011$, and $1400$ required realizations, 
 i.e. better signal to noise but a narrower wave (which will be alleviated by the spreading of Fig.~\ref{fig4}). 
 The latest experiment  \cite{Dubail18} reports $N\simeq4600$,
 $\nu_x\times~\nu_\perp\times\nu_\perp = 8.8\times 7750\times 7750$ Hz and resolution equal to $1.74\mu$m.
 This gives $t=75.91t_c$ ($12.78$ms), $w_{\rm rms}=2.7\mu$m, so looks more favorable for the observation of 1d correlation waves.

Turning to the case of 2d, it is harder to observe the quench. 
Dips of $g^{(2)}$ seen in Fig.~\ref{figg2} on either side of the main peak, 
 will cancel the majority of the highest peak's contribution
 when resolution is poor. Moreover, the peak is lower. For example, 
2d experiments in Chicago  with $^{133}$Cs atoms  \cite{Hung11,Ha13}, %\cite{Chinbook},
had a resolution of 1.8$\mu$m, and a  given interaction strength $g=0.29\frac{\hbar^2}{m}$ and temerature $T=40nK$. 
We calculate a propagation time  of $t=17.05t_c$ (11.25ms), at which one could observe a wave width of $w_{\rm rms}^{(2d)}=1.18\, t^{1/3}=1.7\mu$m and a wave height of $h_{\rm peak}^{(2d)}=0.046 \sqrt{\gamma}/t=0.0007$.

%%%%%%%%%%%%%CONCLUSIONS%%%%%%%%%%%%%%%%%%%%%%%%%%%

 In summary, we have found the universal expressions for the quantum fluctuation contribution to spatial density and
 phase correlations after a quantum quench in dilute Bose gases (\ref{g1int}-\ref{af}), 
 displayed in Figs.~\ref{figg1}~-~\ref{figg2}. The medium and long-time behavior is given
 as simple expressions (\ref{g1sp}-\ref{g2wave}) that are easily applied to assess what can be seen in a given experiment or calculation.
They show how the reduction of phase coherence proceeds after a jump in interaction strength, how antibunching accumulates, and how correlated pairs travel in the gas.
The results apply mostly unchanged to cases in which a quench from $g_0$ to $g\gg g_0$ is made, and to temperatures for which thermal depletion is minor (generally $k_BT\lesssim g\wb{n}$).
Outside of that, thermal effects or those preexisting at $g_0$ may be significant, while our results quantify rather the additional quantum fluctuation contribution induced by the quench.
 Fig.~\ref{fig4} demonstrates that these correlations can be observed even with presently available imaging resolution.
 Conditions for this in 1d experiments \cite{Gring12,Jacqmin11,Armijo11,Dubail18} look realistic. 
 Previously, counterpropagating atom pairs were observed with momentum measurements after expansion such as \cite{Bucker11,Dall09,Perrin07,Wu18}. 
 Looking instead at spatial correlations allows one to observe the different physics of counterpropagating atom pairs \emph{in situ}. 

%%%%%%%%%%%%%ACK%%%%%%%%%%%%%%%%%%%%%%%%%%%%%%%%

\begin{acknowledgments}
We are grateful to 
Pasquale Calabrese,
Peter Drummond,
Tomasz \'Swis\l ocki,
Thomas Gasenzer,
Jan Zill, 
and
Mi\l osz Panfil
 for helpful discussions.
This research was supported by the Marie Curie European Reintegration Grant PERG06-GA-2009-256291, %within the 7th European Community Framework Programme
the Polish Government project 1697/7PRUE/2010/7 and the National Science Centre (Poland) grant No. 2012/07/E/ST2/01389.
\end{acknowledgments}

\bibliography{stobnew}

\end{document}